\documentclass[aps,prd,twocolumn]{revtex4}
\usepackage{graphicx}
\usepackage{bm}

\newcommand{\be}{\begin{equation}}
\newcommand{\ee}{\end{equation}}
\newcommand{\ba}{\begin{eqnarray}}
\newcommand{\ea}{\end{eqnarray}}

\def\aprge{\buildrel > \over {_{\sim}}}

\begin{document}

\title{Hadrons as Signature of Black Hole Production at the LHC}

\author{Irina Mocioiu}
\author{Yasushi Nara}
\author{Ina Sarcevic}
\affiliation{Department of Physics, University of Arizona, Tucson, AZ 85721}

\date{\today}

\begin{abstract}
In models with several large extra dimensions and fundamental 
Planck scale of the order of 1 TeV, black holes can be produced 
in large numbers at LHC energies. We compute the charged hadron 
spectra obtained from the decay of black holes created in $pp$ 
and Pb+Pb collisions at LHC. 
We show that hadrons from black hole decay dominate
at transverse momenta $p_T\aprge 30-100$ GeV/c compared to usual QCD
processes and black hole production signals are easy to identify in 
hadron transverse momentum spectra. Furthermore we show that a measurement 
of the charged hadron spectra probes Planck scales up to 5~TeV for 
any number of extra dimensions.

\end{abstract}

\maketitle

\section{Introduction}

$ $

In scenarios with large extra dimensions the fundamental Planck scale 
could be in the TeV range~\cite{ADD}. One of the most striking consequences 
of a low fundamental Planck scale is the possibility of producing black holes 
and observing them in future colliders or in cosmic rays/neutrino 
interactions~\cite{gtdl, rizzo, bhearly, bhlate}. 

If gravity propagates in $d=4+n$ dimensions while the other fields are 
confined to a 3-brane, the 4-dimensional Planck scale is given by 
$M_{Pl}^2=M_P^{n+2}V_n=G_{n+4}^{-1} V_n$, where $M_P$ is the fundamental 
Planck scale in $4+n$ dimensions and $V_n=(2\pi R)^n$ is the volume of 
the n-torus that describes the compact space. For large size of the extra
dimensions $R$, the fundamental 
scale $M_P$ can be as low as in the TeV range. The existence of large compact 
dimensions leads to deviations from Newtonian gravity at distances of the 
order of R, as well as  strong effects of Kaluza-Klein excitations of the 
graviton on various processes at high energies. These effects impose 
constraints on the scale $M_P$, depending on the number of extra dimensions.
For $n<4$ the strongest limits are given by astrophysical processes
like supernovae cooling and neutron star heating: $M_P\aprge 1500$ TeV for 
$n=2$ and $M_P\aprge 100$ TeV for $n=3$. Cosmological considerations give 
similar bounds for $n=2$ and $n=3$ and they imply $M_P\aprge 1.5$ TeV even 
for $n=4$. These bounds contain, however, a larger degree of uncertainty 
and model-dependence. Non-observation of black hole production in cosmic 
ray interactions also imposes constraints for $n\ge 4$ at the level 
of $M_P\aprge 1$ TeV. Present collider limits are typically 
below 1 TeV for any number of large extra dimensions. 

Given all these constraints, we will concentrate here on the scenario
with $n=6$ and $M_P\sim 1-5$ TeV. We will also 
discuss the dependence of our results on $n$ and $M_P$. One thing to note is 
that the constraints mentioned above are derived for a toroidal 
compactification where all the large extra dimensions have the same radius. 
For different types of compactification the limits could actually be 
considerably relaxed.

Black hole production and evaporation can be described semiclassically and 
statistically when the mass of the black hole is very large compared to the 
fundamental Planck mass. When the mass of the black hole approaches $M_P$ one
expects quantum gravity effects to become important. We want to 
explore only the parameter space where the semi-classical treatment is 
justified. The total available energy at LHC is 14 TeV.
Black holes with masses of this order can be produced in $pp$ collisions. 
For Pb+Pb collisions, the black holes produced would have masses up to 5.5 TeV.
These masses should be high enough above the Planck scales considered here 
for the semi-classical description to be valid~\cite{gtdl}.

In Ref.~\cite{supress},
 it was suggested that the geometrical cross-section would be 
exponentially suppressed. Detailed subsequent 
studies~\cite{nosuppress},\cite{YN} 
did not confirm this proposal and showed that the geometrical cross-section is 
modified only by a numerical factor of order one.
 In Ref.~\cite{rizzo}, it was shown that
 even including the exponential suppression of the cross-section,
the production rates are still high. We will use here the geometrical 
cross-section. 

These black holes decay very rapidly. The decay occurs in several stages.
For the purpose of detecting black hole events, the most important phase is
the semi-classical Hawking evaporation, since it provides a large multiplicity 
of particles and a characteristic black-body type spectrum. Most of this 
Hawking radiation is on the brane~\cite{ehm}, producing all Standard
Model particles. Because most of the Standard Model degrees of freedom come 
from strongly interacting particles (quarks and gluons), hadrons will be the 
dominant signal for the events where black holes are formed. 

In this paper we compute the transverse momentum distributions
 of charged hadrons at mid-rapidity 
obtained from the evaporation of black holes produced in $pp$ collisions 
and Pb+Pb collisions at LHC energies.
 We show that in $pp$ collisions the black hole 
events produce a large number of hadrons and dominate over the 
QCD background at transverse momenta above around 30-100 GeV/c,
 where they can be clearly measured. The results have a 
weak dependence on the number of large extra dimensions, but depend 
quite strongly on $M_P$. The signal is big enough to detect even for
 $M_P\sim 5$ TeV. For Pb+Pb collisions the energy available is lower,
so one can only produce lower mass black holes and probe lower Plank scales. 
However, the rates could be higher due to the large number of 
binary collisions. We also discuss some of the possible 
uncertainties that would affect our results.

\section{Black Hole Production and Decay}

In a high energy parton-parton collision,
 the impact parameter could be smaller 
than the Schwarzschild radius in $d$ dimensions for a black hole with 
mass $M_{BH}$:
\be
r_h=\frac{1}{\sqrt{\pi}}\frac{1}{M_P}\Bigg[\frac{M_{BH}}{M_P}\Bigg(\frac{8
\Gamma(\frac{n+3}{2})}{n+2}\Bigg)\Bigg]^{\frac{1}{n+1}}\ .
\ee
This leads to the formation of a semi-classical $d$-dimensional black hole of 
size $r_h$ much bigger than the fundamental Planck scale $M_P$, but much 
smaller than the size of the large extra dimensions $R$.

With the above assumptions, the black hole production in a parton-parton 
collision is given by the geometrical cross-section $\sigma_{BH}=\pi r_h^2$.
Then the cross-section in a $pp$ collision is obtained by
folding in the parton densities:
\ba
\sigma(pp&\to& BH+X) ={1 \over s}
\sum_{a,b}\int^{{s}}_{M^2_{BH,min}}dM_{BH}^2 \nonumber\\
& \times & \int^1_{x_{1,min}} 
 {dx_1\over x_1}
 f_a(x_1,Q^2)\sigma_{BH}f_b(x_2,Q^2)
\label{prod}
\ea
where $x_1$ and $x_2={M_{BH}^2/( x_1 s)}$ are the momentum fractions of the 
initial partons and $x_{1,min}=M^2_{BH}/s$. 
The scale in the parton distribution functions  $f(x,Q^2)$ is chosen 
to be $Q^2=1/r_h^2$. The results depend only weakly on the choice of this 
scale. We use CTEQ6M~\cite{cteq6} for the parton distribution functions.

The radiation rate into Standard Model particles is given by a thermal 
distribution in 4 dimensions:
\begin{equation}
  {dE\over dt}= {1\over (2\pi)^3}\sum_{i} \int \frac{\omega  g_i\sigma_{i,s} 
d^3k}
     {e^{\omega/T_{BH}}\pm 1}
\label{rate}
 \end{equation}
with the black hole temperature:
 \begin{equation}
 T_{BH}= {d-3 \over 4\pi r_{h}},
\end{equation}
 where the sum is over all Standard Model particles and $g_i$ is a 
statistical factor, counting the number of degrees of freedom. The sign in 
the denominator is $+$ for fermions and $-$ for bosons.
$\sigma_{i,s}$ are the gray body factors, which depend on the spin $s$ of each
particle. We first approximate these by  
$\sigma_{i,s} = \Gamma_s A_4$, where $\Gamma_s$ are constant \cite{page}
($\Gamma_{1/2}=2/3,\Gamma_{1}=1/4, \Gamma_0=1 $). 
A black hole acts as an absorber with a radius somewhat larger than 
$r_h$, such that $A_k$ can be written as \cite{ehm}:
\begin{equation}
 A_k = \Omega_{k-2}\left( {d-1 \over 2} \right)^{{d-2\over d-3}}
 \left( {d-1\over d-3} \right)^{{k-2\over 2}} r_{h}^{k-2}
\end{equation}
with
\begin{equation}
 \Omega_k= {2\pi^{k+1\over 2} \over \Gamma({k+1\over 2})}.
\end{equation}

We compare these results with those obtained by using the recently computed 
gray body factors given in eq. (47) and (53) of Ref.~\cite{gray}.
The differences in the final results are found to be around 30\%.

For the emission into gravitons, which are $d$-dimensional, the rate is given by:
\be
   {dE\over dt}= {1\over (2\pi)^{d-1}}\sum_{i} \int \frac{\omega  g_i
\sigma_i d^{d-1}k}
   {e^{\omega/T_{BH}}- 1}\ ,
 \ee
where $\sigma_i\sim A_d$. This will be much smaller than the emission 
rate into SM particles \cite{ehm} and we can neglect it when calculating 
the lifetime.

The lifetime of the black hole is then obtained by integrating eq.(\ref{rate}).
Assuming no mass evolution during the decay,
we get:
\ba
\tau_{BH}&=&M_{BH}\Bigg[ {\pi^2\over 30} \Big(\sum_{f} {7\over 8} g_{f} 
\sigma_{f} +
\sum_b g_b \sigma_b\Big) T_{BH}^4\Bigg]^{-1}\nonumber \\
&=& c M_{BH} {1\over r_h^2 T_{BH}^4}= c' {1\over M_P} \Big({M_{BH}\over M_P}
\Big)^{n+3\over n+1}\ .
\ea

\section{Hadron spectrum}

We compute the cross section for inclusive charged hadron production
from the partons produced in the black hole decay from
\begin{eqnarray}
 E{d\sigma^h\over d^3p} &=& {1\over s}\sum_{a,b,c}\int^{{s}}_{M_{BH,min}^2}dM_{BH}^2
\int^1_{x_{1,min}} {dx_1\over x_1} \int_{z_{min}}^{1}{dz\over z^2} 
\nonumber \\
&\times& f_a(x_1,Q^2)\sigma_{BH}f_b(x_2,Q^2)
E_c{dN_c\over d^3p_c} \nonumber\\
&\times& D_c^h(z,Q_f^2),
\label{spectrum}
\end{eqnarray}
where $z_{min}=2p/\sqrt{s}$ and $z=p/p_c$ and the decay distribution is:
\begin{equation}
  E_c{dN\over
    d^3p_c}= {1\over (2\pi)^3} {p_c^{\mu}u_{\mu}\gamma  g_c\sigma_c
     \tau_{BH}
    \over e^{p_c^{\mu}u_{\mu}/T_{BH}}\pm 1},
\end{equation}
where $\gamma$ is the Lorentz gamma factor and 
$u=(\gamma,0,0,(x_1-x_2)\sqrt{s}/(2 M))$ takes into account that the
black hole is not produced at rest, but with a small velocity. Here $p$ and 
$E$ refer to hadrons, while $p_c$ and $E_c$ are for partons. 

We choose the scale of the fragmentation function $D_c^h(z,Q_f^2)$ to be
the final transverse momentum $p_T$ of the hadrons. 
The KKP fragmentation function~\cite{kkp} is used to get the final charged 
hadrons from the partons produced in the evaporation of the black hole.
The KKP fragmentation function is only parametrized in the
range of $0.1\leq z \leq 1.0$ and $1.4 \leq Q_f \leq 100$ GeV.
We need to access  small values of the transverse momentum fraction
$z=p_{hadron}/p_{parton}$, as well as large $Q_f$, 
for the partons from black hole decay.
For the large $Q_f$, we explicitly evolve the KKP fragmentation
function from the scale $Q_f=100$ GeV up to the desired values
(in this case up to $\sim 10$ TeV) using DGLAP equations~\cite{qcdnum}.
For the $z < 0.1$ range,
we use small-$z$ fragmentation function by Fong and Webber~\cite{FW91}
which is based on the coherent parton branching formalism, which
correctly takes into account the leading and next-to-leading
soft gluon singularities, as well as the leading collinear ones.
It was found that the predicted energy dependence
of the peak in the $\xi = \ln(1/z)$ distribution agrees very well
with the $e^+e^-$ annihilation data up to c.m. energy of 200 GeV
\cite{TASSO90}.

\begin{figure}[t]
\includegraphics[width=3.3in]{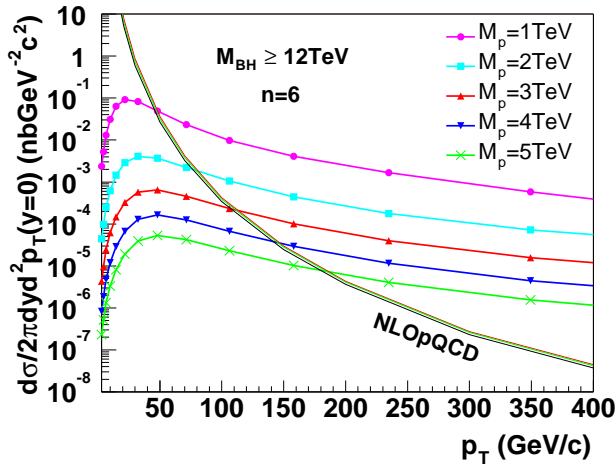}
\caption{Cross section for inclusive charged hadron
 production from 10-dimensional 
black holes in $pp$ collision at $\sqrt{s}=14$ TeV
compared to the QCD background as a function of $M_P$.
Black hole masses are integrated over
 $ 12\rm{TeV} \leq M_{BH} \leq \sqrt{s}$.
NLOpQCD predictions are plotted for the scales $Q=Q_f= 2p_T,p_T, p_T/2$.
}\label{ppmp}
\end{figure}

\begin{figure}[ht]
\includegraphics[width=3.3in]{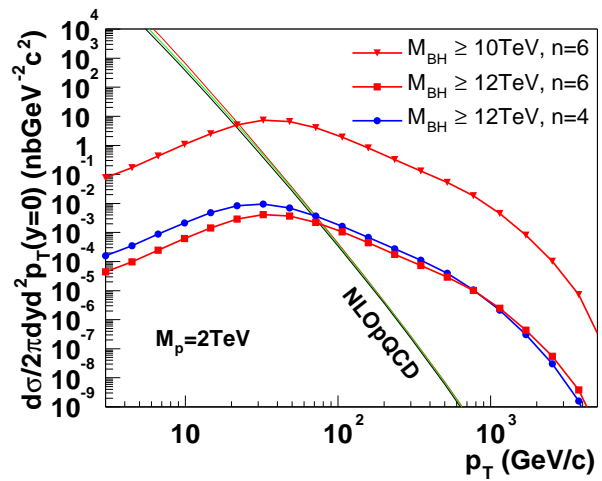}
\caption{Cross section for inclusive charged hadron production
 from black holes in $pp$ collision at $\sqrt{s}=14$ TeV for $n=4$ and 
$n=6$ and different minimum masses of the black holes produced.
}\label{pp}
\end{figure}

Fig.~\ref{ppmp} shows the cross section for inclusive charged hadron 
production from black holes for several values of $M_P$ ranging from 
1 to 5 TeV in $pp$ collision at LHC,
 compared to the expected spectrum of hadrons from QCD.
LHC will be sensitive enough to detect the QCD hadrons up to $p_T$ 
around 400 GeV/c. The black hole signal is much bigger than the QCD one 
starting 
at $p_T\sim 50-200$ GeV/c, depending on the Planck scale. It can be seen that 
even for $M_P$ as high as 5 TeV there is a 
considerable signal above background at $p_T\aprge 200$ GeV/c. 
At higher $p_T$ the background is practically inexistent, while the black 
hole signal is still very large, as seen in Fig.~\ref{pp}. 

We show in Fig.~\ref{pp} the cross-section for inclusive charged hadron 
production 
from black hole decay for $M_P=2$ TeV, compared with the QCD background. 
It is reasonable to assume that black holes with masses only slightly higher 
than $M_P$ can be produced. However,  
as previously discussed, our semiclassical description of the black holes is 
only valid for black hole masses much bigger than the fundamental Plank scale. 
Integrating over the mass of the black hole starting at a low minimum value 
(close to $M_P$) would assume the validity of this treatment beyond its 
range of applicability. We choose to integrate starting at a much higher 
black hole mass. Clearly, black holes with masses lower than
our cut-off can be produced, but results obtained with a cut-off at $M_P$ 
would have to take into account quantum gravity effects which are unknown. 
We show here results for 
$M_{BH}^{min}=10$ TeV and $M_{BH}^{min}=12$ TeV. It can be seen that including 
lower mass black holes gives considerably higher rates. Consequently, we 
consider our approach to be a `conservative' one: our results are an 
underestimate of the actual signal and our qualitative conclusions always 
hold, while the actual quantitative results could be much higher than our 
estimates, making the signal  easier to detect. Even for high mass
black holes the signal clearly dominates over the QCD background in the region 
above 100 GeV/c, where it can be easily seen in the experiments. Including 
lower mass black holes gives a bigger signal for all momenta and also drives 
the signal above the background even for lower momenta, of the order of tens 
of GeV. 

We also study the dependence of the results on the number of extra dimensions 
and show that it is very small, as can be seen in Fig~\ref{pp}. 

The QCD background \cite{qcd} 
is shown for different choices of the scale used in the 
structure and fragmentation functions (we use $p_T, p_T/2, 2p_T$).
The dependence on this scale is very weak in the high transverse momentum 
region. For the black hole signal, the same change in 
$Q_f$ leads to differences of up to a factor of 2 in the results, which 
does not change any of our conclusions.

We notice that even though there are significant changes in the overall rate
of hadrons produced, the transverse momentum dependence of
the hadrons does not change 
much when changing $M_P$ or $M_{BH}^{min}$. This is not the case at the parton
level. Changing $M_P$ or $M_{BH}^{min}$ the temperature of the black hole is
modified and consequently the spectrum of the emitted particles is different.
Even though we can still see this for partons, the hadronization washes out 
most of the effect. We conclude that we cannot get a direct determination of 
the temperature of the black holes from the hadron spectrum. One could attempt 
to do that by looking at the spectrum of photons and electrons in the black 
hole event, which preserves the black body radiation type of 
spectrum~\cite{gtdl}, but is considerably lower than the hadron signal 
because photons 
and electrons are only a small fraction of the particles produced in the black 
hole evaporation. In that case, one would be forced to consider black holes 
with lower masses in order to obtain a detectable signal.

We conclude from here that black hole events will be easily detected in 
the hadron spectra at high $p_T$. The values of $p_T$ for which the signal
becomes higher than the background would give an indication on the values
of $M_P$ and $M_{BH}$ that this signal corresponds to. 

\section {Black Holes in Pb+Pb Collisions}

To compute the spectra in the case of Pb+Pb collisions,
we multiply the expression in 
Eq.~(\ref{prod}) by the Glauber profile density
$T_{AA}({b})=\int d^2r T_A({r}) T_A(|\bm{b}-\bm{r}|)$, 
where $T_A({r})=\int dz \rho({r},z)$, 
normalized such that $\int d^2r T_A({r})=A$,
$A$ being the atomic mass number and $\rho({r},z)$
the nuclear density for which we take the Woods-Saxon distribution.
This factor gives an enhancement for the production cross-section. 
For example, $T_{AA}(b=2{\rm fm})=28$ mb$^{-1}$
in the case of Pb+Pb collisions at impact parameter $b=2$ fm.
However, $\sqrt{s_{NN}}=5.5$ TeV in this case, so only lower mass black holes 
can be 
produced and a smaller parameter space can be probed.
We do not include shadowing effects, since black holes dominate at high $x$, 
where these are negligible. 
In addition, it is demonstrated that nuclear modification
of parton distribution is getting smaller when we go to
larger scale from $Q^2=2.25$ GeV$^2$ up to $Q^2=(100 
{\rm GeV})^2$~\cite{eks98}.
We have confirmed that this also holds true for much higher
scales up to $Q=30$ TeV by evolving the nuclear parton density in \cite{eks98} 
and conclude they are negligible.

\begin{figure}[ht]
\includegraphics[width=3.3in]{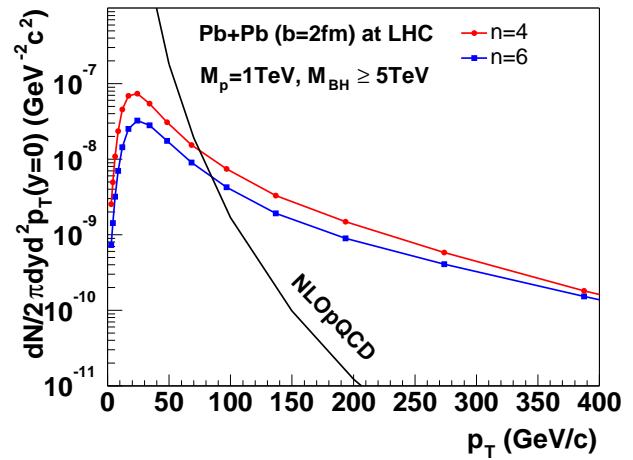}
\caption{Transverse momentum distribution for charged particle
at mid-rapidity from black hole decay in Pb+Pb collision
at impact parameter $b=2$ fm  at LHC.
NLOpQCD calculation with the scale $Q=Q_f=p_T$ is also shown.
}\label{fig:pbpb}
\end{figure}

In Fig.~\ref{fig:pbpb} we show the results for Pb+Pb collisions 
for $M_P=1$ TeV and
$M_{BH}^{min}=5$ TeV. For these parameters, the signal is still significant
and can be easily detected. Due to the fact that $\sqrt{s_{NN}}$ is 
only 5.5 TeV, the higher scales are no longer accessible in 
this type of experiment.


In Pb+Pb collisions,
the black hole is expected to be produced in a dense medium of quarks and 
gluons,
therefore we need to take into account the interactions of the 
partons produced in the decay of the black hole with the
quarks and gluons around it, for example, as in Ref.~\cite{eloss}.
The energy loss is expected to be small at high transverse momentum.
At LHC energies, for $p_T\sim 100$ GeV/c, where the black hole signal 
clearly dominates over the background, the energy loss was found to be small, 
(about 5\% effect)~\cite{vitev}.

 However, energy loss has significant
effects at $p_T$ below 10 GeV in the QCD spectrum at LHC~\cite{vitev}.
For the hadrons coming from black holes we also expect the effect to be 
small in the high momentum region.
However, there is a possibility to have {\it enhanced}
particle yield around $p_T \sim 10$ GeV/c, because the 
hadron spectra is much flatter than that of the QCD spectra
and feedback from the emitted gluons could be non-negligible.
If this is the case, the black hole signal could be also identified in the 
lower $p_T$ region, in addition to the high $p_T$ one. 
It could happen that the signal becomes higher than the background at 
these values of $p_T$. Even if the signal is somewhat lower than 
the background, it is still 
large, such that the experiment would detect a large number of hadrons from 
signal+background, even at $p_T$'s of tens of GeV.

We do not include interactions of the 
black hole itself with the surrounding particles. One can imagine taking 
into account possible absorption of these particles by the black hole, 
which would affect the decay of the black hole. This is an interesting 
issue and is presently under investigation \cite{prep}.

\section{Conclusions}

In summary, we have computed the transverse momentum spectra
for high $p_T$ charged hadrons from decay of black holes produced in 
$pp$ collisions at $\sqrt{s}=14$ TeV
as well as
in central Pb+Pb collisions at $\sqrt{s_{NN}}=5.5$ TeV.
We have shown that the hadrons from black holes are detectable and dominate 
the background for $p_T$ above about 100 GeV for fundamental Planck scales 
up to 5 TeV. Our results are conservative, as they only take into account 
very high mass black holes. Including black holes with lower masses gives 
even stronger signals. The value of $p_T$  at which the signal becomes bigger 
than the background is determined by $M_P$ and $M_{BH}$ considered.

We have neglected the evolution of the mass of the black hole 
during the decay. If we take into account that, as the mass 
decreases, the temperature increases, we would get a somewhat harder 
spectrum. In the same time, the lifetime would decrease compared 
to our estimate, so that our curves would move slightly down and to the right.
However, all the qualitative features previously discussed will remain the 
same. 
The last stage of the decay, when the mass of the black hole has decreased to 
almost $M_P$ is not understood, since it requires a full quantum gravity 
description. This is why a full consideration of the mass evolution during 
decay is not really possible.
We have not taken into account the angular momentum of the
black hole and the phase of the decay when the black hole would just loose
this angular momentum. Also, there are a few additional particles produced in
the initial stage of the black hole decay, when the black hole looses
the quantum numbers of the partons that produced it.
In \cite{YN} it has been shown that classical gravitational radiation could be
important and consequently the actual black hole mass is smaller than that 
of the center of mass energy of the parton collision. This would reduce
the number of very high mass black holes produced. However, even in case of 
 a small fraction of initial energy going into black hole production, as 
long as $M_{BH} \gg M_P$,  
there is still a large, 
observable signal since production of black holes with small mass is large.  

 If there are additional degrees of freedom around 100 GeV (new particles), so 
below $T_{BH}$, they should be taken into account and they would lead to a 
small decrease in the lifetime. Their contribution would be slightly 
suppressed, just as for the top quark, W's, Z's and Higgs, for which the 
masses are no longer much smaller than the temperature. 
These issues introduce some uncertainty in the numerical results, 
but our main conclusions are unaffected.

We would like to note that the QCD background we show in the graphs is 
computed at $y=0$. For high rapidity this background is 
actually much smaller, while the black hole signal is the same for all 
rapidities. This would indicate that by looking in the high rapidity region 
one would enhance the signal to background ratio even further.

In conclusion, we have shown that the charged hadron distribution in
$pp$ and Pb+Pb collisions at LHC energies provides an unique probe of
black hole production and the physics of extra dimensions for
Planck scale up to 5 TeV and for any number of extra dimensions.

\section{Acknowledgments}
We thank Radu~Roiban for useful discussions. We would also like to 
thank P.~Aurenche and J.~P.~Guillet for providing us the code for
calculating NLOpQCD hadron distributions. 
This work has been supported in part by the DOE under Contracts
DE-FG02-95ER40906 and DE-FG03-93ER40792.

\end{document}